\documentstyle[aps,preprint]{revtex}
\begin{document}
\draft
\preprint{submitted to  Z. Physik B}

\title{ RKKY interaction in Layered Superconductors with Anisotropic
Pairing}
\author{D.N.Aristov, S.V.Maleyev, A.G.Yashenkin}

\address{Petersburg Nuclear Physics Institute\\
Gatchina, St.Petersburg 188350, Russia}
\date{Received \hspace{4cm} .}
\maketitle

\begin{abstract}
The RKKY interaction between rare-earth  (RE) ions in high-$T_c$
superconductors is considered at $T\ll T_c$. It is shown that this
interaction consists of two terms: conventional oscillating one and
the positive term, which is proportional to the gap function and
decreases in the $2D$ case inversely proportional to the distance. In
the antiferromagnetic state of the RE subsystem this positive
interaction gives rise for frustrations which diminishes the Neel
temperature. In the case of strongly anisotropic gap function this
frustration produces two different values of the effective nearest
neighbor exchange coupling between RE ions along the $a$ and $b$.
This anisotropy has been established experimentally in
Ref.\cite{6,7,8}.
\end{abstract}



In many High-$T_c$ superconductors substitution of some ions by
rare-\-earth elements (RE) leads to the low-\-temperature
antiferromagnetism in the RE subsystem  with $T_N$ of order of
1~ K and without sizeable change of the superconducting
transition temperature $T_c$. The mostly investigated compounds
are YBa$_2$Cu$_3$O$_{7-x}$ $(1:2:3)$ and YBa$_2$Cu$_4$O$_8$
$(1:2:4)$ with yttrium being substituted by different RE ions
(see \cite{1,2,3,4,5} and references therein).

The magnetic dipolar, superexchange and RKKY interactions has
been attracted for explanation of this antiferromagnetism in the
RE subsystem. The detailed calculations, however,  have
been carried out only for dipolar interaction. Corresponding
results have been presented in \cite{4} and \cite{5} for
$(1:2:3)$ and $(1:2:4)$ compounds, respectively. In particular,
it was pointed out in \cite{4} that there is no correlation
between the ground state energies calculated in the dipolar
approximation and observed Neel temperatures. Hence the
dipolar interaction alone cannot explain the magnetic
properties of the considered systems. Moreover, a striking
feature of the magnetic interaction between the RE ions is the
very large difference between effective nearest neighbor
exchange interactions along crystallographic $a$ and $b$
directions.  Indeed,
the
specific heat data near $T_N$ for $1:2:3$ system with Sm and Nd
ions was very well fitted
 \cite{6,7}
using $2D$ Ising model with
strongly different exchange constants $J_1$ and $J_2$.
The ratio $J_1/J_2$ was found  to be  $\simeq 11$ and 50 for Sm
and Nd samples respectively \cite{6}.  Recent measurements for
NdBa$_2$Cu$_3$O$_{7-\delta}$ \cite{7} confirmed this result  and
revealed that the ratio $J_1/J_2$ depends on doping and
increases rapidly with $\delta$. As it was emphasized in
Ref.\cite{7} such a  behaviour indicates that chains cannot be
the direct reason of this anisotropy, since the structural
anisotropy decreases with depleting of the chains.

More direct confirmation of this anisotropy has been provided
\cite{8} by the inelastic neutron scattering in
HoBa$_2$Cu$_3$O$_{7\delta}$.  The values of
$J_a=(0.0005\pm0.0002)$meV and $J_b=-(0.0028\pm0.0008)$meV were
reported there.

The orthorombic distortion in the considered compounds is very small
$[(a-b)/a\sim10^{-2}-10^{-3}]$ and weakly depends on $\delta$.
Therefore, it cannot  explain the anisotropy of superexchange and
dipolar  interactions. At the same time it is well known that the
hole system  responsible for the superconductivity possesses
pronounced anisotropy in the a-b plane which strongly depends on the
oxygen content $\delta$. This anisotropy was established   for the
Fermi surface \cite{9,10} and is apparently presented in the
superconducting gap function $\Delta_{{\bf k}}$ \cite{10,11}.
Moreover, the anisotropy of $\Delta_{{\bf k}}$ should persist in any
system with anisotropic Fermi surface. As a result we may suggest
that the above exchange anisotropy of the RE subsystem
stems from the interaction between the RE ions and holes in CuO$_2$
planes.

To date we have no direct information about  interaction between
the RE ions and CuO$_2$ planes in metal state. Meanwhile for
antiferromagnetic compound HoBa$_2$Cu$_3$O$_{6.13}$ this interaction
has been evaluated on the base of experimentally observed
splitting of the Holmium $\Gamma_5$ doublet. This splitting
appears due to the interaction of  Ho ion with spinwaves in
CuO$_2$ planes \cite{12}. It was found that exchange integral
between Ho total angular momentum and Cu$^{2+}$ spin is equal to
2.4 meV.  Therefore we see that this interaction is not so weak
and the same should be valid for the metal state, as well.

The RE-hole interaction gives rise to the RKKY interaction
between the RE ions, which should be anisotropic due to the
anisotropies of the Fermi surface and of the gap function,
mentioned above. In this paper we consider the problem of
anisotropy of the RKKY interaction in superconducting state
which stems from the anisotropy of the gap function. The effects
arising from the non-\-circular shape of Fermi surface will be
considered elsewhere.

We obtain the following principal result. At $T\ll T_c$ the $2D$ RKKY
interaction is a sum of two terms. The first one is the conventional
oscillating contribution and the second term is positive and
decreases as $R^{-1}$. It is proportional to $|\Delta_{{\bf R}}|/E_F$
where $\Delta_{{\bf R}}$ is the gap function along ${\bf R}$
direction.  Both terms are screened at distances of order of
$V_F/|\Delta_{{\bf R}}|$. The second term should strongly
frustrate the AF ground state due to its long-\-range behavior.
If the gap $\Delta_{{\bf R}}$ has not square symmetry this
frustration is different in the $a$ and $b$ directions. As a
result the effective exchange parameters $J_a$ and $J_b$ should
be different, too.

This paper is organized as follows. In Sec.II we evaluate the
$2D$ RKKY interaction for $k_FR\gg1$ in the case of anisotropic
gap function and in Sec.III we discuss the physical consequences
of these results.

\section{RKKY interaction between RE ions in superconductors with
anisotropic pairing}

We begin with the conventional form of the exchange interaction
between RE total angular moment ${\bf J}$ and the hole spin
density $1/2\,{\bf\sigma}({\bf r})$ in CuO$_2$ planes \cite{13}
     \begin{equation}
     V({\bf r})=-J_{ex}(g_J-1)
     {\bf J}({\bf r}) {\bf\sigma}({\bf r})\ ,
     \end{equation}
where $J_{ex}$ is the exchange interaction and $g_J$  is the
ionic $g$-\-factor. As a result we may write the RKKY
interaction between two RE ions in the following form
     \begin{equation}
     H_{RKKY}=2[J_{ex}(g_J-1)]^2 {\bf J}_1{\bf J}_2
     I({\bf R}_{12})\ ,
     \end{equation}
where below $T_c$ we have
      \begin{equation}
     I({\bf R}) =  T\sum_n \left\{
     G(i\omega_n,{\bf R})^2 +
     |F(i\omega_n,{\bf R})|^2     \right\}\ .
     \label{defRKKY}
     \end{equation}
Here $\omega_n=\pi T(2n+1)$ is the Matsubara frequency and the
normal and anomalous Green functions are given by
     \begin{equation}
     G(i\omega_n,{\bf R}) =
     \frac{a^2}{(2\pi)^2} \int d^2{\bf k} \exp(i {\bf kR})
     \frac{
     i\omega_n + \xi_k }{\omega_n^2+|\Delta_k|^2+
     \xi_k^2}
     \label{defG-R}\end{equation}

     \begin{equation}
     F(i\omega_n,{\bf R}) =
     \frac{a^2}{(2\pi)^2} \int d^2{\bf k} \exp(i {\bf kR})
     \frac{ \Delta_k }
     {\omega_n^2+|\Delta_k|^2+ \xi_k^2}
     \label{defF-R}\end{equation}
where $a^2$ is the volume of the square unit cell of the $2D$
lattice, $\xi_{{\bf k}}$ is the hole spectrum above $T_c$ and
$\Delta_{{\bf k}}$ is the superconducting gap function.
The conventional form of writing of the RKKY interaction below
$T_c$ \cite{14} may be obtained from eqs. (2)--(5) after
summation over $\omega_n$.

We are interested in the dependence of $I({\bf R})$ on the
direction of ${\bf R}$ in the $ab$ plane. The case of a
non-\-circular shape of the Fermi surface appears to be a quite
complex problem and will be analyzed elsewhere. Remarkable fact
presented below is the possibility to obtain the expression for
the RKKY interaction at large distances for the circular Fermi
surface and arbitrary form of the gap function $\Delta_{{\bf
k}}$. Hence we take the following parametrization of the
normal-\-state dispersion $\xi_{{\bf k}}$ and the function
$\Delta_{{\bf k}}$:
     \begin{equation}
     \xi_{{\bf k}}=\frac{k^2-k^2_F}{2m};
     \qquad
     \Delta_{{\bf k}}=\Delta_\varphi\ ,
     \end{equation}
where $\varphi$ is the polar angle in the $ab$ plane.

We begin with evaluation of asymptotic expressions for $G$ and $F$
functions at $k_FR\gg1$ and then determine the corresponding form of
the RKKY interaction. Taking into account that
$E_F\gg\Delta_\varphi,T$ we may put $k=k_F+\xi/V_F$ where
$V_F=k_F/m$. Then, integrating in Eq.(4) over $\xi$, we get
     \begin{eqnarray}
     G({\bf R},\omega)&=&
     \frac{a^2m}{2\pi i}\int d\varphi e^{ik_FR\cos\varphi}
     \frac{\omega+\sqrt{\omega^2+
     \Delta^2_{\varphi+\psi}}\,sign(\cos\varphi) }
     {2\sqrt{\omega^2+\Delta^2_{\varphi+\psi}}}
     \\
     &\times&
     \exp\left[-\frac R{V_F}
     \sqrt{\omega^2+\Delta^2_{\varphi+\psi}}|\cos\varphi|\right],
     \nonumber
     \end{eqnarray}
where  $\psi$
defines the direction of ${\bf R}$. The expression for
$F({\bf R},\omega)$ is obtained, if one replaces
$(\omega\pm\sqrt{\omega^2+\Delta^2_{\varphi+\psi}})$ by
$i\Delta_{\varphi+\psi}$ in (7).

If $k_FR\gg1$, the steepest decent method may be used. Expanding
$\cos\varphi$ near the points $\varphi=0,\pi$ and taking
into account that $\Delta^2_{\varphi}=\Delta^2_{\varphi+\pi}$ we
have
     \begin{eqnarray}
     G({\bf R},\omega)&=&
     \frac{a^2m}{(2\pi k_FR)^{1/2}}
     \left[\frac{-i\omega}{\sqrt{\omega^2+\Delta^2_{{\bf R}}}}
     \cos\left(k_FR-\frac\pi4\right)\right.
     \nonumber\\
     &&+\left.\sin\left(k_FR-\frac\pi4\right)\right]
     \exp\left[-\frac R{V_F}\left(\omega^2+
     \Delta^2_{{\bf R}}\right)^{1/2}\right]
     \\
     F({\bf R},\omega)
     &=&\frac{a^2m}{(2\pi k_FR)^{1/2}}
     \frac{\Delta_{{\bf R}}}
     {\sqrt{\omega^2+\Delta^2_{{\bf R}}}}
     \cos\left(k_FR-\frac\pi4\right)
     \exp\left[-\frac R{V_F}(\omega^2+
     \Delta^2_{{\bf R}})^{1/2}\right],
     \end{eqnarray}
where $\Delta_{{\bf R}}$ stands for the absolute value
$|\Delta_\varphi|$ at $\varphi= \psi$. We drop the phase of
$\Delta_\varphi$ in the numerator of (9), since it is irrelevant
for the interaction (3).

Expressions (8) and (9) take place if one can neglect the
variation of $\Delta_{\varphi+\psi}$ in the range
$|\varphi  +\psi |\lesssim(2/k_FR)^{1/2}$ at $R\lesssim
V_F(\Delta^2_{{\bf R}}+\omega^2)^{-1/2}$. These conditions may
be combined in the following form
     \begin{equation}
     \frac{(\Delta_{{\bf R}}\Delta'_{{\bf
     R}})^2}{E_F(\omega^2+\Delta_{{\bf
     R}}^2)^{3/2}}
     \sim\frac{\Delta'^2}{E_R\Delta_{{\bf R}}}\ll1\ .
     \end{equation}
Here in the right-hand side we put $\omega\sim
T\lesssim\Delta_{{\bf R}}$. For most directions we have
$\Delta'_{{\bf R}}\sim\Delta_{{\bf R}}$ and the condition (10)
fulfills very well since always $\Delta_{{\bf R}}\ll E_F$.
However it is violated if the gap function has a node at
$\varphi=\varphi_n$.  Indeed, in this case
we have $\Delta_{{\bf R}}=\Delta_n|\psi-\varphi_n|$ near the node
and instead of (10) we get $\Delta_n/[E_F|\psi-\varphi_n|]\ll1$.
Therefore near the directions of nodes the Eqs. (8) and (9)
persist if
     \begin{equation}
     |\psi-\varphi_n| >|\Delta_n|/E_F\ .
     \end{equation}
For example, in the case of the $d_{x^2-y^2}$
pairing we may write approximately $\Delta_\varphi\simeq\Delta_0
\cos2\varphi$. In this case we have $\varphi_n=\pm\pi/4$;
$\pm3\pi/4$ and $\Delta_n=\pm2\Delta_0$.

From the condition (11) we see that Eqs.(8) and (9) hold almost in
all directions of ${\bf R}$ excluding narrow vicinities the gap
nodes. We begin with the regions where condition (11) fulfills. Using
Eqs.(3),(8) and (9) we get

     \begin{eqnarray}
     I({\bf R})&=&
     -\frac{a^4m}{2\pi k_FR}T\sum_{\omega_n}
     \left[
     \frac{\omega^2_n}{\omega^2_n+\Delta^2_{{\bf R}}}
     \sin2k_FR- \frac{\Delta^2_{{\bf R}}}
     {\omega^2_n+\Delta^2_{{\bf R}}}\right]
     \nonumber\\
     &\times& \exp\left[-\frac{2R}{V_F}
     (\omega^2+\Delta^2_{{\bf R}})^{1/2} \right]\ .
     \end{eqnarray}
We are interested in very low temperatures of order of 1~ K or
less.  In this case we may put $T=0$ and instead of (12) we have
     \begin{eqnarray}
     I({\bf R})&=&
     -\frac{a^4m^2}{2\pi^2k_FR}\int^\infty_0d\omega\left(
     \frac{\omega^2}{\omega^2+\Delta^2_{{\bf R}}}
     \sin2k_FR\right.\nonumber\\
     &-&\left.\frac{\Delta^2_{{\bf R}}}
     {\omega^2+\Delta^2_{{\bf R}}}\right)
     \exp\left[-\frac{2R}{V_F}(\omega^2+
     \Delta^2_{{\bf R}})^{1/2}\right].
     \end{eqnarray}
In the normal state when $\Delta_{{\bf R}}\equiv0$
we  obtain from this equation the asymptotic form of the
two-\-dimensional RKKY interaction

     \begin{equation}
     I_0({\bf R})=-\frac{a^4m}{4\pi^2R^2}\sin2k_FR\ .
     \end{equation}

For $\Delta_{{\bf R}}\neq0$ after some algebra we get

     \begin{eqnarray}
     I({{\bf R}})&=&
     \frac{a^4mk_F^2}{4\pi^2}
     \left[-\frac{\sin2k_FR}{(k_FR)^2}F_1
     \left(\frac{2R\Delta_{{\bf R}}}{V_F}\right)
     \right.\nonumber \\
     &+&\left. \frac{\Delta_{{\bf R}}}
     {E_F(k_FR)}F_2\left(
     \frac{2R\Delta_{{\bf R}}}{V_F}\right)\right]\ ,
     \end{eqnarray}
where
     \begin{eqnarray}
     F_1(x)=\int\limits^\infty_x\frac{dye^{-y}}y(y^2-x^2)^{1/2}=&1; &
     x\ll1 \nonumber\\
     &\left(\frac\pi{2x}\right)^{1/2}e^{-x};&x\gg1
     \nonumber\\
     F_2(x)=\int\limits^\infty_x\frac{dye^{-y}}y
     \frac x{(y^2-x^2)^{1/2}}=
     &\frac\pi2;
     &  x\ll1 \\
     &\left(\frac\pi{2x}\right)^{1/2}e^{-x};& x\gg1 .
     \nonumber
     \end{eqnarray}

There are two important features of these equations. First, the
superconducting gap gives rise to the screening of the RKKY
interaction. The range of the screening depends strongly on the ${\bf
R}$ direction and is given by
\begin{equation}
\xi_{{\bf R}}=\xi_m \frac{\Delta_m}{\Delta_{{\bf R}}} \ ,
\end{equation}
where $\xi_m=V_F/\Delta_m$ and $\Delta_m=\max(\Delta_{{\bf R}})$.
Note here that the value of $\xi_m$, apparently related to
$T_c$, is usually adopted as the superconducting correlation
length.  We see that it is {\em the lowest} value of the
screening length for the case of anisotropic pairing.

Second, in addition to the usual oscillating term, the
long-\-range positive interaction appears. This interaction is
proportional to $|\Delta_{{\bf R}}|/E_F$ and depends strongly on
direction of ${\bf R}$.  In spite of its weakness, this part of
the interaction should be important due to its very slow
decreasing with $R$, if $R<\xi_{{\bf R}}$. In the vicinity of
the AF transition it should produce strong frustration and
diminish the Neel temperature.

We considered above the  directions of ${\bf R}$ determined by
the condition (11).  In the opposite limiting case when
$|\varphi-\varphi_n| \ll|\Delta_n|/E_F$, the screening range
(17) seemingly goes to infinity. We will show now that it is not
the case.

If we put $\psi = \varphi_n$ in Eq.(7), we
have $\Delta_{\varphi+\psi}=\Delta_n \varphi$. As a result the
$\Delta^2_\varphi$ contribution to the exponential factor in (7)
becomes important at $R>(E_F/\Delta_n)^2/k_F$. Thus the range of
screening along the nodes is given by

     \begin{equation}
     \xi_n= (E_F/\Delta_n)^2/k_F \sim
     \xi_m(E_F/\Delta_n)
     \gg\xi_m\ .
     \end{equation}


At the same time similarly to Eq.(7) the expression for $F({\bf
R},\omega)$ is zero due to the oddness of the integrand as a
function of $\varphi$ in this case.

Summarizing we can say that Eq.(15) is a good approximation for
$I({\bf R})$ at $k_FR\gg1$ if near the nodal directions one
replaces the argument of function $F_1$ by
$2R\Delta^2_n/V_FE_F$.

\section{Discussion}

We have demonstrated that below $T_c$ RKKY interaction given by
Eq.(15) is a sum of two terms. First is conventional one
which in the $2D$ case oscillates as $\sin2k_FR$ and decreases
as $(k_FR)^{-2}$. Second term is strongly anisotropic positive
(of the antiferromagnetic sign) contribution proportional to the
small parameter $\Delta_{{\bf R}}/E_F$ and decreasing as
$(k_FR)^{-1}$.  Both terms are screened at distances determined
by Eqs. (17) and (18). However these equations hold in the pure
case only. In real systems interaction with impurities may  lead
to a lower value of the screening range, especially in the
directions of nodes.

In spite of a small factor $\Delta_{{\bf R}}/E_F$ the second term in
Eq.(15) is important due to its very slow decreasing with $R$.
In presence of long-\-range antiferromagnetic order in the
Rare-\-Earth subsystem it produces frustration and reduces the
Neel temperature.

Let us return now to the mentioned in Introduction  anisotropy
of the exchange parameters in the $a$--$b$ plane. In \cite{6,7,8}
the specific-\-heat and neutron scattering data were described
using $2D$ anisotropic exchange in $a$--$b$ plane. Of course, it is
only a convenient parametrization and the real interaction is
much more complex. However an anisotropy in the $a$--$b$ plane is
one of its most important features. The above mentioned positive
part of the RKKY interaction may be one of sources of such
anisotropy, if the gap function has not  square symmetry, e.g.
if it is a mixture of $d_{x^2-y^2}$ and extended $s$-wave
function as discussed in \cite{10}. In this case
$\Delta_{{\bf b}}\neq\Delta_{{\bf a}}$ and corresponding
frustration becomes different in these two directions.
Effectively this situation may be modelled by nearest-\-neighbour
exchange parameters $J_a\neq J_b$ as it was actually done in
Refs.\cite{6,7,8}.

It should be noted, however, that the observed asymmetry of the Fermi
surface in $a$--$b$ plane is  another possible source of
anisotropy of the RKKY interaction.

\acknowledgements
We thank A.Buzdin, A.Furrer, K.Kikoin, M.Kiselev, I.Vagner
for useful discussions.
One of us (D.N.A.) thanks LNS at ETH\&PSI for the hospitality.
This work was partly supported by International Science Foundation
and Russian Goverment (Grant \#R3Y300) and Russian Foundation for
Basic Researches (Project No.~93-02-2224).

\end{document}